\documentclass{ws-ijmpd}

\usepackage[super,compress]{cite}
\usepackage[dvips,breaklinks]{hyperref}
\hypersetup{colorlinks,urlcolor=black,citecolor=black,linkcolor=black,filecolor=black}
\usepackage{breakurl}
\usepackage[super,compress]{cite}
\begin{document}

\markboth{Kamble \& Kaplan}
{EM Counterparts of GW Sources : Mergers of Compact Objects}

%
\catchline{}{}{}{}{}
%
\title{ELECTROMAGNETIC COUNTERPARTS OF GRAVITATIONAL WAVE SOURCES : MERGERS OF COMPACT OBJECTS}


\author{ATISH KAMBLE\footnote{University of Wisconsin-Milwaukee}}
\address{Harvard-Smithsonian Center for Astrophysics, 60 Garden St.,
\\ Cambridge, MA 02138, USA
\footnote{akamble@cfa.harvard.edu}}

\author{DAVID L. A. KAPLAN}

\address{Physics Department, University of Wisconsin-Milwaukee,\\
1900 E. Kenwood Blvd, \\Milwaukee, WI 53211, USA
\footnote{kaplan@uwm.edu}}

\maketitle

\begin{history}
\received{Day Month Year}
\revised{Day Month Year}
\end{history}

\begin{abstract}
Mergers of compact objects are considered prime sources of  
gravitational waves (GW) and will soon be targets 
of GW observatories such as the Advanced-LIGO, VIRGO etc.
Finding electromagnetic counterparts of these GW sources 
will be important to understand their nature.
We discuss possible electromagnetic signatures of the mergers.
We show that the BH-BH mergers could have luminosities which exceed
Eddington luminosity from unity to several orders of magnitude
depending on the masses of the merging BHs.
As a result these mergers could be explosive, release up to $10^{51}$ erg
of energy and shine as radio transients. At any given time we expect 
about a few such transients in the sky at GHz frequencies
which could be detected out to about 300 Mpc.
It has also been argued that these radio transients would 
look alike radio supernovae with comparable detection rates.
Multi-band follow up could, however, distinguish between the mergers 
and supernovae.
\end{abstract}


\ccode{PACS numbers:}

\section{Introduction}

Compact binary mergers, such as neutron star-neutron star (NS-NS),
neutron star-black hole (NS-BH) and black hole-black hole (BH-BH)
are of great interest to gravitational wave (GW) astronomers because of the possibility 
of detecting GWs from nearby such events in the near future by using 
advanced sensitivity of GW observatories. While some of the ground based 
GW detectors are already operational to designed sensitivities and 
some are being upgraded, several space based GW missions are in
the initial stages of design studies. For example Laser Interferometric 
Gravitational wave Observatory (LIGO) and Virgo are being upgraded 
to second generation of GW detectors to be called advanced-LIGO and 
advanced-Virgo, respectively. Among the space based detectors 
Laser Interferometer Space Antenna (LISA)
and DECI-Hertz Interferometer Gravitational wave Observator (DECIGO)
are at different levels of design studies. Space based detectors are mostly 
sensitive to GW events in the frequency band $10^{-4}-0.1$ Hz and 
ground based detectors to those in the range $1-10^{4}$ Hz.
At the even lower end of the frequency range will be 
Astrodynamical Space Test of Relativity using Optical Devices (ASTROD-GW)
\cite{WTN2008} space based mission whose design is being studied and 
which will operate between $10^{-7}-10^{-3}$ Hz. 

GWs from the inspiralling compact binaries will probe Einstein's General 
Relativity as well as alternative theories of gravity to an unprecedented accuracy
in the strong field regime 
(see reviews by Arun K. G. and Kent Yagi in this issue). Detecting EM counter-part 
to GW signals is of immense importance for fundamental physics.
Coincident EM-GW signals could potentially be used to build independent
cosmological distance ladder \cite{Bloom2009,Holz2005}, to probe
the central engines of supernovae and gamma ray bursts (GRBs) 
as well as to study the evolution of massive stars and formation
of compact objects such as BHs and NSs, to study galaxy-super massive 
black hole (SMBH) symbiosis and evolution \cite{Schnittman2011}
and even to measure neutron star equation of state \cite{Faber2012}.
Therefore it is also important to know what could be the electromagnetic
signal produced by these mergers.

It has been proposed that the nucleosynthesis of heavy radio-active elements 
during the NS-NS merger could heat the ejecta and power a transient
which will shine in optical/near-IR bands \cite{Metzger2010,Metzger2012}. 
This mechanism is similar to the one that powers optical brightness 
of supernovae. These transients would have luminosities somewhere between 
novae and supernovae with absolute visual magnitude $M_{v} \approx -15$ 
and would reach peak brightness on a timescale of $\approx$ 1 day
and could be detected out to $\approx Gpc$ distances.

Numerical simulations have made great progress in the last few years
and have been providing deeper understanding of the issues involved in 
mergers and of the processes such as accretion and radiation during 
the short span of mergers. In short, these simulations predict that mergers of
two compact objects, such as NS-NS $\&$ BH-BH mergers, could produce outflows
with high velocities ($\beta \sim0.1-0.8$), release energies up to
$10^{49}$\,erg \cite{2005ApJ...634.1202R,Nakar2011,Piran2012,
Palenzuela2010} and could power expansion of blast-waves in the surrounding medium.
Such blast-waves would inevitably produce radio signals which could be detected
from up to 300 Mpc or farther by the existing and upcoming radio
telescopes.

Several astrophysical phenomena produce high velocity blast waves
e.g. jets in AGNs and micro quasars, GRBs, supernovae etc. The shock
waves generated during these explosions are the primary sources of
associated radio transients in the universe. Their radio brightness is
powered chiefly by synchrotron emission generated by the
relativistic shock-heated electrons gyrating in the post-shock
amplified magnetic fields. 
While majority of these events are discovered through the relatively
short lived high-energy emission (optical, X-ray or $\gamma$-ray)
associated with the explosion itself some events are likely to be
discovered more efficiently by the radio emission. Radio observations
of a recent type Ib/c supernova, SN2009bb, showed that it was
expanding at mildly relativistic velocities ($\beta \approx 0.8$)
powered by the central engine \cite{2010Natur.463..513S}. 
Such SN are usually associated with
GRBs and are discovered by their intense but short $\gamma$-ray energy
emission. Yet, SN2009bb had no detected $\gamma$-ray counter-part.

It is not clear if the mergers of compact objects will have associated prompt emission 
at high energies. However, we show here that a survey of radio sky at $\approx$ GHz 
waveband would be ideal to discover radio transients produced by compact object
mergers. With the advent of modern radio telescopes having deeper sensitivity 
and wide-field imaging capabilities it will be possible in near future to carry 
out such blind surveys to discover transients associated with the GW sources.
Multi-wavelength observations will be useful in early
identification of the nature of these sources.

\section{Energy release during the merger of compact objects}\label{sec:energy}
A rotating black hole has a reservoir of free energy associated with its 
spin angular momentum which, in principle, could be extracted
in various ways resulting in slowing down the rotation of the black hole.
Ref.\refcite{Penrose1969} demonstrated this theoretically by carefully choosing 
particle trajectories; this particular energy extraction process is now known as the Penrose process.
Subsequently, scattering of vacuum electromagnetic \cite{Press1972} or
gravitational \cite{Hawking1972,Teukolsky1974} or magnetohydrodynamic (MHD) waves
\cite{Uchida1997} were also shown to be possible means of tapping 
the energy. However, the Blandford-Znajek \cite{BZ1977} (BZ) mechanism 
of extracting the energy electromagnetically via Poynting flux appears 
to be the most likely workable mechanism in astrophysical settings. 
Numerical simulations have further shown the BZ mechanism not only 
to be viable but also a good candidate
for powering variety of astrophysical sources such as the relativistic jets 
in active galactic nuclei \cite{}, x-ray binaries \cite{} and gamma ray 
bursts \cite{Lee2000}. A significant amount of energy could be released 
in the process with luminosity 
$\rm L \sim 3 \times 10^{43} ~B^{2}_{4}~M^{2}_{8} ~erg~s^{-1}$ where
B is the magnetic field strength in Gauss through which the BHs of M solar mass moves. 
Throughout this paper we will follow the notation $\rm Q_{a} = Q/10^{a}$.

Several authors have carried out numerical simulations of BH-NS \& NS-NS mergers using 
either Newtonian or General Relativistic dynamics and to a varying degree 
of detailed treatment of microphysics. Despite these differences all simulations agree 
to the conclusion that significant amount of energy is released in fast moving ejecta  
during the mergers. For example, in the NS-NS mergers Ref.~\refcite{Rosswog1999} find
$10^{51}$ erg of energy released to the ejecta moving with $\beta = 0.2$
where $\beta$ is the ejecta velocity in units of speed of light. All NS-NS mergers
lead to formation of accretion disk which opens up possibilities for other sources
of energy e.g. neutrino-antineutrino annihilation, BZ mechanisms etc. and possibly 
relativistic outflows. Overall, in the NS-NS merger, the amount of energy released 
could vary between $10^{49}-10^{52}$ erg and outflow velocities from non-relativistic 
to relativistic.

On the other hand simulations of BH-NS mergers produce mildly relativistic but energetic 
ejecta with $\beta=0.5$ and $10^{52}$ erg. BH-NS merger rates are quite 
uncertain and therefore their detectability is difficult to predict. But given their high 
energies and ejecta velocities they could comprise a non-negligible fraction of 
the overall gravitational and electromagnetic wave detections.

Recent simulations of SMBH mergers in the presence of imposed magnetic
field suggested energy release in the form of Poynting flux outflow \cite{Mosta2010,
Palenzuela2010}. Ref.~\refcite{Palenzuela2010} simulated merger of two $10^{8} M_{\odot}$
BHs. The resulting Poynting flux flare had a luminosity of 
$\rm L\approx 4\times 10^{43} ~erg ~s^{-1}$ and lasted for about $\rm 5 ~hr$ \cite{Neilsen2011}
with total energy release of $\approx 2 \times 10^{48} erg$ in Poynting flux.

\section{Radiation and dynamics of the outflows: Spectra and light curves}
Consider an outflow with isotropic energy released in the explosion 
being $E$. In the case of supernovae typical observed energies are 
about $10^{46}-10^{47}$\,erg while those for compact mergers are 
expected to be about $10^{49}$\,erg. The explosion drives a shock wave 
expanding into the surrounding medium which we consider to have 
a homogeneous density profile. The shock-wave sweeps the surrounding
medium and heats it to relativistic temperatures. The shock converts bulk
kinetic energy of the incoming material into thermal energy of the shocked material.
The electrons in the shock-heated plasma are accelerated in the post-shock 
magnetic field to radiate synchrotron radiation.  Below we derive the
expected flux density evolution.

\subsection{The Electron energy distribution}
We adopt the conventional power law Lorentz factor distribution of
shocked electrons with power-law index $p$:
\begin{equation}
n_{e}(\gamma_{e})~ d\gamma_{e} = K_{e}~ \gamma_{e}^{-p}~ d\gamma_{e}
\label{eqn:e_distribution}
\end{equation}
for $\gamma_{m} \ll \gamma_{e} \ll \gamma_{u}$, where $\gamma_{m}$ and $\gamma_{u}$
are the lower and upper energy cut-offs of the distribution and $\gamma_{e}$ the electron
Lorentz factors. 

At this point it is important to note that not all the shocked electrons will end
up in the non-thermal energy distribution. This fraction depends on the speed of the shock-wave.
When the shock-wave is ultra-relativistic most of the shocked electrons can end up building 
the non-thermal distribution, which may be the case for GRBs. It may not be true, however,
for non-relativistic blast waves such as in the present case and in the supernovae.
While a detailed treatment of the dependence of $\epsilon_{r}$ on the shock speed is beyond 
this article, for simplicity we assume that fraction to be a constant.
The value of $\gamma_{m}$ and $K_{e}$ can then be found be assuming that a constant 
fraction $\epsilon_{r}$ of all  the shocked electrons end up in the non-thermal electron energy 
distribution and that the amount of energy available for the electron acceleration is a constant
fraction ($\epsilon_{e}$) of the total thermal energy density $U_{th}'$.
\begin{eqnarray}
\int_{\gamma_{m}}^{\gamma_{u}} n_{e}(\gamma_{e})~d\gamma_{e} &=& 4 \epsilon_{r} n \label{eqn:e_conserve} \label{eqn:e_density}\\
\int_{\gamma_{m}}^{\gamma_{u}} \gamma_{e} ~m_{e} c^{2} ~n_{e}(\gamma_{e})~d\gamma_{e} &=& \epsilon_{e} U'_{th}
\label{eqn:e_energy}
\end{eqnarray}
The primed quantities are measured in their local rest frame i.e. in
the frame of the shocked material. Shock wave efficiently converts its kinetic energy 
into internal energy of the shocked material. For a shock wave moving several times faster 
than the local sound speed density compression of factor 4 is achieved and 
the compressed medium trails the shock wave with speed lower by that factor.
Further, it can be shown that the thermal energy density of the shocked medium 
is given by $U_{th} = (9/8) n m_{p} \beta^{2} c^{2}$.  
Using Eqn. \ref{eqn:e_distribution}, \ref{eqn:e_density},
\ref{eqn:e_energy} and assuming $\gamma_{u} \gg \gamma_{m}$ one
obtains
\begin{eqnarray}
\gamma_{m} & = & \epsilon_{e} \frac{9}{32}\frac{m_{p}}{m_{e}} ~\frac{p-2}{p-1}  \beta^{2} \label{eqn:gamma_m}\\
K_{e} & = &4 \epsilon_{r} n(p-1)\gamma_{m}^{p-1}    \label{eqn:Ke}
\end{eqnarray}

\subsection{Post-shock magnetic field}

It is assumed that similar to $\epsilon_{e}$ a fraction $\epsilon_{B}$
of the post-shock thermal energy goes into the magnetic field.
\begin{eqnarray}
\frac{B'^{2}}{8\pi} &=& \epsilon_{B} U'_{th}
\label{eqn:B_at_rest}
\end{eqnarray}
Similar to $\epsilon_{e}$, in our entire discussion we will treat $\epsilon_{B}$
as a constant in time.

\subsection{Synchrotron Spectrum}
It is assumed that the shocked electrons gyrate in the post-shock magnetic fields 
and radiate synchrotron radiation. 
Ref.~\refcite{1999ApJ...523..177W} generalised the standard synchrotron
function of Ref.~\refcite{1986rpa..book.....R} by integrating over isotropic
distribution of pitch angles and then estimated the dimensionless
frequency $x_{p}$ at which the dimensionless flux peaks $F(x_{p}) = \phi_{p}$. 
These values, they show, are considerably different than
the usually used standard values. Therefore, we will continue to
follow the treatment and notation of Ref.~\refcite{1999ApJ...523..177W}.
For a power law distribution of radiating electrons, the values of $x_{p}$
and $\phi_{p}$ depend on the distribution index $p$ as shown in 
Ref.~\refcite{1999ApJ...523..177W}, where
for our fiducial $p=2.5$ we get $x_p=0.5$ and $\phi_p=0.65$.
We adopt the shape of the synchrotron spectrum 
as given by Ref.~\refcite{1998ApJ...497L..17S}.

\subsubsection{$\gamma_{m}$ and corresponding spectral break in the Synchrotron Spectrum}
The power-law distribution of the electrons has a lower Lorentz factor cut-off
which we identify as $\gamma_{m}$ (Equation~\ref{eqn:gamma_m}).
We adopt expression by Ref.~\refcite{1999ApJ...523..177W} for the characteristic synchrotron 
frequency corresponding to $\gamma_{m}$,

\begin{equation}
\nu_{m}' = \frac{3x_{p}}{4\pi}\,\frac{\gamma_{m}^{2} q B'}{m_{e} c},
\label{eqn:nu_m_at_rest}
\end{equation}
where $q$ is the electric charge, $m_{e}$ is the mass of electron and
$x_{p}$ is the dimensionless frequency discussed in \cite{1999ApJ...523..177W}. 
The value of $x_{p}$ depends on the electron energy
distribution index $p$.  The spectrum will fall for
$\nu^\prime>\nu_m^\prime$.

\subsubsection{Spectral break due to synchrotron self absorption}
By approximating the thickness of the shocked radiating plasma to be $dr = r/12$ 
the optical depth can be approximated as $\tau_{\nu} \sim \alpha_{\nu} ~dr$. 
For the synchrotron self-absorption coefficient we used equation 6.53 of Ref.~\refcite{1986rpa..book.....R}
and inverted the relation $\tau_{\nu} = 1.0$ to obtain $\nu_{a}'$.

\subsubsection{Peak flux of the Synchrotron Spectrum}
Assuming that a fraction $\epsilon_{r}$ of all the shocked electrons, $N_e$, contribute to 
the non-thermal synchrotron radiation, the instantaneous peak flux at $\nu_{m}$ can be estimated as
\begin{equation}
F_m=F_{\nu}(\nu_m) =  \frac{\epsilon_{r} N_{e} P'_{\nu_{m}}}{4 \pi d_{L}^{2}} \label{eqn:Fm}
\end{equation}
where $N_{e} = (4 \pi /3) R^{3} n (1+X)/2$ and $P'_{\nu_{m}}$ is the synchrotron power 
per electron averaged over an isotropic distribution of pitch angles \cite{1999ApJ...523..177W}
\begin{equation}
P'_{\nu_{m}} = \phi_{p} \frac{\sqrt{3}q^{3} B'}{m_{e} c^{2}}.
\label{eqn:power_per_electron}
\end{equation}
We also have the luminosity distance $d_L$ and the hydrogen fraction $X$.

\subsection{Evolution of the Spectrum}
The density profile of the surrounding medium which the shock-wave is ploughing
through plays an important role in dictating the dynamics of its evolution and
subsequently that of the radiation spectrum. There is a strong evidence that
massive stars are the progenitors of radio SNe of Type Ib/c \cite{}. This means 
that the SN generated shock-wave should be expanding into the stellar wind 
of the progenitor star. Compact objects which are products of SNe such as NSs 
receive significant kick velocity at the birth and are thrown out of the location 
where they were born. Therefore, it is expected that the merger of compact 
objects would take place in the homogeneous inter-stellar medium or in 
the distant galactic halos. 
Considering this we calculate the spectral and temporal evolution
in homogeneous ISM below in section~\ref{sec:ISM}. For completeness
we calculate the evolution in wind density profile also in section~\ref{sec:wind}.
 
\subsubsection{Homogeneous inter-stellar medium}\label{sec:ISM}
The instantaneous synchrotron spectrum can then be characterized 
by two break frequencies ($\nu_{a}'$ and $\nu_{m}'$) and the peak flux $F_{{m}}'$.
Because the shock wave, and therefore the radiating plasma behind it, is expected
to be moving at non-relativistic or mildly relativistic velocities we have
neglected relativistic effects including doppler boosting, for simplicity.
As a result of this we use $\nu_{m}' = \nu_{m}$, $\nu_{a}' = \nu_{a}$
and $F_{{m}}' = F_{{m}}$ where unprimed quantities are   
measured at an observer on the Earth.
For the temporal evolution, initially the ejecta will coast along with
a constant velocity $\beta (=\beta_{\rm dec})$ and therefore $R =
\beta_{\rm dec} c t$. 
Equations \ref{eqn:gamma_m}-\ref{eqn:power_per_electron} then give
\begin{eqnarray}
F_{{m}}(t_{\oplus,d}) 	&	\approx	&	0.15(\gamma_{dec}\beta_{dec})^4
n^{3/2}t_{\oplus,d}^3 (X+1) (z+1) \sqrt{\epsilon_B} \epsilon_{r} \phi_p
d^{-2}_{L,100}\,\mu{\rm Jy}\\
\nu_{m}(t_{\oplus,d})		&	\approx	&
10^{12}(\gamma_{dec}\beta_{dec})^5\sqrt{n}(p-2)^2x_{p}\sqrt{\epsilon_B}\epsilon_{e}^2
(p-1)^{-2}(X+1)^{-2} \,{\rm Hz}
\end{eqnarray}
along with the synchrotron self absorption frequency
\begin{equation}
\nu_{a}(t_{\oplus,d}) =  10^7 n^{4/5} (p-1)^{8/5} t_{\oplus,d}^{3/5}(X+1)\epsilon_B^{1/5}(\gamma_{dec}\beta_{dec})^{-1}(p-2)^{-1}~ \left(\frac{p+\frac{2}{3}}{p+2}\right)^{-3/5}\epsilon_{e}^{-1}\epsilon_{r}^{3/5} .
\end{equation}
when $\nu_{a} \ll \nu_{m}$ and 
\begin{equation}
\nu_{a}(t_{\oplus,d}) =  10^{12}
n^{\frac{p+6}{2(p+4)}}
t_{\oplus,d}^{\frac{2}{p+4}}
(1+X)^{\frac{-2(p-1)}{p+4}}
\epsilon_B^{\frac{p+2}{2(p+4)}}
\epsilon_{e}^{\frac{2(p-1)}{p+4}}
\epsilon_{r}^{\frac{2}{p+4}}
(\gamma_{dec}\beta_{dec})^{\frac{5p}{p+4}}
\Gamma(\frac{3p+2}{12})
\Gamma(\frac{3p+22}{12})
\end{equation}
when $\nu_{a} \gg \nu_{m}$. We take $t_{\oplus,d}$ to be the time since the explosion as measured
by the observer in days and $d_{L,100}$ to be the luminosity distance in units of 100 Mpc.

After the ejecta has swept up enough matter, comparable to
its kinetic energy, it will start to decelerate. This will happen
at a distance
\begin{equation}
R_{\rm dec} \approx 10^{17} E_{49}^{1/3}  n^{-1/3} \beta_{\rm
  dec}^{-2/3}\,{\rm cm}
\end{equation}
and at time
\begin{equation}
t_{\rm dec} \approx 45 E_{49}^{1/3} n^{-1/3} \beta_{\rm
  dec}^{-5/3}\,{\rm days}
\end{equation}
After this epoch shocked plasma would assume Sedov-von Neumann-Taylor
(SNT) self similarity and the ejecta will decelerate with $\beta =
\beta_{\rm dec} (R/R_{\rm dec})^{-3/2}$ while its temporal evolution is given
by $R = R_{\rm dec} (t/t_{\rm dec})^{2/5}$. Therefore the evolution of the spectrum
for $t>t_{dec}$ is given by :
\begin{eqnarray}
F_{{m}}(t) 	&	=	&	F_{{m}}(t_{\rm dec}) \left(\frac{t}{t_{\rm dec}}\right)^{3/5} \\
\nu_{m}(t)		&	=	&      \nu_{m} (t_{\rm dec}) \left(\frac{t}{t_{\rm dec}}\right)^{-3} \\
\nu_{a}(t)		&	=	&      \nu_{a} (t_{\rm dec}) \left(\frac{t}{t_{\rm dec}}\right)^{6/5}
\end{eqnarray}

\subsubsection{Wind inter-stellar medium}\label{sec:wind}
During their lifetime massive stars drive strong winds and loose mass. In the process 
density profile of the immediate environment around these stars gets modified. 
For a constant rate of mass loss $\dot{M}$, and constant wind velocity $V_{w}$ 
the interstellar medium density around the star assumes a power law profile $\rho = A r^{-2}$ up to 
the wind termination shock. By assuming $\dot{M} = 10^{-5} {\rm M_{\odot} ~yr^{-1}}$
and $V_{w} = 1000 {\rm~km ~s^{-1}}$, $A$ can also be written down 
as $A = \dot{M}/(4 \pi V_{w}) = 5 \times 10^{11} A_{\star} {\rm ~gm~cm^{-1}}$.
Using the wind density profile along with Equations 
\ref{eqn:gamma_m}-\ref{eqn:power_per_electron} and the fact that initially 
the ejecta will be coasting along at a constant velocity $\beta (=\beta_{\rm dec})$ 
and therefore $R = \beta_{\rm dec} c t$ gives
\begin{eqnarray}
F_{{m}}(t_{\oplus,d}) 	&	\approx	& 	1.4 \times 10^{6} A_{\star}^{3/2} (1+X) (1+z) (\gamma_{dec}\beta_{dec})  \sqrt{\epsilon_{B}} \epsilon_{r} \phi_{p}
   d_{L,100}^{-2} ~\mu Jy\\
\nu_{m}(t_{\oplus,d})	&	\approx	& 	2 \times 10^{14} \sqrt{A_{\star}} \left( \frac{p-2}{p-1}\right)^2 x_{p} (\gamma_{dec}\beta_{dec})^4 \sqrt{\epsilon_{B}} \epsilon_{e}^2 t_{\oplus,d}^{-1} (1+X)^{-2} ~{\rm Hz}\\
\nu_{a}(t_{\oplus,d})	&	\approx	& 	8 \times 10^{10} A_{\star}^{4/5} \left[ \frac{ (p-1)^{8/5} (p+2)^{3/5}} {(p-2) (3 p+2)^{3/5}} \right] (1+X) \epsilon_{B}^{1/5} t_{\oplus,d}^{-1}  (\gamma_{dec}\beta_{dec})^{-13/5} \epsilon_{e}^{-1} \epsilon^{3/5}_{r}~{\rm Hz}
\end{eqnarray}

This evolution will continue until the shock-wave sweeps material comparable
to its kinetic energy after which the ejecta will decelerate. This will happen at
a distance of $R = R_{dec}$ given by
\begin{equation}
R_{\rm dec} \approx 5.3 \times 10^{15} E_{49} A_{\star}^{-1} \beta_{dec}^{-2} \,{\rm cm}
\end{equation}
and at time $t_{dec} = R_{dec}/\beta_{dec}c$ :
\begin{equation}
t_{\rm dec} \approx 2 E_{49} A_{\star}^{-1} \beta_{dec}^{-3} \,{\rm days}
\end{equation}
After this epoch the shock wave expands as $R = R_{\rm dec} (t/t_{\rm dec})^{2/3}$
and the ejecta decelerates as $\beta =\beta_{\rm dec} (R/R_{\rm dec})^{-1/2}$.
The evolution of the spectrum for $t>t_{dec}$ is given by :
\begin{eqnarray}
F_{{m}}(t) 	&	=	&	F_{{m}}(t_{\rm dec}) \left(\frac{t}{t_{\rm dec}}\right)^{-1/3} \\
\nu_{m}(t)		&	=	&      \nu_{m} (t_{\rm dec}) \left(\frac{t}{t_{\rm dec}}\right)^{-7/3} \\
\nu_{a}(t)		&	=	&      \nu_{a} (t_{\rm dec}) \left(\frac{t}{t_{\rm dec}}\right)^{-2/15}
\end{eqnarray}
The evolution of all the spectral parameters thus obtained is summarized and 
compared with that in the homogeneous density profile in Table
~\ref{tab:ta1}.  

\begin{table}[ph]
\tbl{{\bf Time Dependence of Spectral Parameters : }
We give the temporal power-law index (i.e., $a$ in $x(t)\propto t^a$
for parameter $x$) of the peak flux density $F_{{m}}$, the
peak frequency $\nu_m$, and the self-absorption frequency $\nu_a$
(considered in two limits compared to $\nu_m$).  Here, $p$ is the
power-law index for the electron energy distribution: $n_e(\gamma_e)
\propto \gamma_e^{-p}$.}
{\begin{tabular}{@{}ccccc@{}} \toprule
				& \multicolumn{4}{c}{Time Dependence} \\ \colrule
Parameter& \multicolumn{2}{c}{Homogeneous density profile} & \multicolumn{2}{c}{Wind density profile}\\
 				 		& $t \ll t_{dec}$	&	$t \gg t_{dec}$		& $t \ll t_{dec}$&$t \gg t_{dec}$\\ \colrule
$F_{{m}}$\dotfill			&	3		&	3/5				&	0		&	-1/3	\\
$\nu_{m}$\dotfill			&	0 		& 	$-3$ 			&	-1		&	-7/3	\\
$\nu_{a} (\ll \nu_{m})$\dotfill	&	3/5 		& 	6/5 				&	-1		&	-2/15	\\
$\nu_{a} (\gg \nu_{m})$\dotfill	&	$2/(p+4)$	&	${-(3p-2)/(p+4)}$	&	-1		&	-(7p-6)/3(p+4)	\\ \botrule
\end{tabular} \label{tab:ta1}}
\end{table}

The overall lightcurve could be calculated using the spectral shape 
and the evolution of spectral breaks in Table ~\ref{tab:ta1}. For the simplest case of optically thin spectrum, 
i.e. $\nu_{a} \ll \nu_{m} \ll \nu$, the light curve after merging the pre- and post-deceleration regimes, is given by:
\begin{equation}
F_{\nu}(t) = F_{\nu}(t_{\rm dec}) \left[
  \left(\frac{t}{t_{\rm dec}}\right)^{-\alpha_{1}s}+ \left(\frac{t}{t_{\rm
      dec}}\right)^{-\alpha_{2} s}\right]^{-1/s},
\label{eqn:lightcurve}
\end{equation}
where $\alpha_{1} = 3.0$ and $\alpha_{2}=3/5+3b=-0.3(5p-7)\approx -5/3$ for
$p=2.5$, and we use a value $s=2$ to smoothly join the two
regimes. The peak of the light curve which occurs at $t = t_{\rm dec}$
is obtained by $F_{\nu}(t_{\rm dec}) = F_{{m}}
\left(\frac{\nu}{\nu_m}\right)^{b}$ where $b=-(p-1)/2=-3/4$ for
$p=2.5$, and by using instantaneous values of $\nu_{m}$ and $F_{{m}}$
at $t=t_{\rm dec}$.  So at a constant observing frequency,
$F_\nu\propto t^3$ for $t<t_{\rm dec}$ and $F_\nu \propto t^{-5/3}$
after deceleration has begun.

\section{Expected Event Rates of NS-NS mergers}
It can be shown that for the reasonable values of parameters the spectrum 
should be optically thin at radio wavelengths with $F_{\nu} \propto \nu^{-(p-1)/2}$ 
by the time they reach their peak brightness around
$t=t_{\rm dec}$. Their peak brightness at a specific observing
frequency could therefore be given as:
\begin{eqnarray}
F_{\nu,\rm peak} &\approx& 23 f_{p}E_{49} n^{(p+1)/4} \beta_{\rm
  dec}^{(5p-7)/2} \epsilon_{B}^{(p+1)/4} \epsilon_{e}^{p-1} \epsilon_{r}
d^{-2}_{L,100}\, \nu_{\rm
  GHz}^{-(p-1)/2}\,{\rm mJy} \nonumber \\
&\approx& 128 E_{49}n^{7/8}\beta_{\rm
  dec}^{11/4}\epsilon_B^{7/8}\epsilon_e^{3/2} \epsilon_{r} d^{-2}_{L,100} \nu_{\rm
  GHz}^{-3/4}\,{\rm mJy}
\end{eqnarray}
where $f_{p} =  \phi_{p} \left(\frac{x_{p}} {3.2 \times 10^{-3}} \right)^{(p-1)/2} \left(\frac{p-2}{p-1}\right)^{p-1} $. 
A radio telescope with a sensitivity of about $50\,\mu{\rm Jy}$
at $\approx $\,GHz frequencies can detect such transients out to about 300\,Mpc.

If $\cal R$ is an intrinsic event rate then an all sky-snapshot should
be able to detect $N_{\rm all-sky} = {\cal R}V\Delta t$ events which occur
within the volume $V$ limited by the telescope sensitivity and remain
detectable for $\Delta t$ amount of time. Following
Ref.~\refcite{Nakar2011,Piran2012} we set $\Delta t \approx t_{\rm dec}$ and
use ${\cal R}_{\rm CM} =300{\cal R}_{300}\,{\rm Gpc}^{-3}\,{\rm yr}^{-1}$ for NS-NS merger rate
giving
\begin{equation} 
N_{\rm all-sky} \approx 1 E_{49}^{11/6} n^{47/48} \beta_{\rm
  dec}^{59/24} \epsilon_{B,{0.1}}^{21/16} \epsilon_{e,{0.1}}^{9/4} \epsilon_{r,0.3}^{3/2}
{\cal R}_{300}\left(\frac{F_{\nu,\rm lim}}{50\,{\mu\rm Jy}}\right)^{-3/2}\nu_{\rm GHz}^{-9/8}
\end{equation}

It is clear from above equation that the detection rate is strongly
dependent on two parameters: the energy released in the blast-wave and
its velocity. 
For the reasons of simplicity we
have used the approximation $\Delta t \approx t_{\rm dec}$ where
$\Delta t$ is the duration for which the event remains above the
telescope sensitivity and therefore is visible. In practice, $\Delta t
> t_{\rm dec}$ by a factor of few and therefore actual rate of detection 
could be higher by that factor.

\section{BH-BH binary mergers}
It has been shown that black hole's huge reservoir of energy, in the form of its
rotational energy, could be tapped at least theoretically via the Blandford-Znajek 
mechanism. Some of the brightest and energetic objects in the universe have 
been thought to be powered by this mechanism which plays central role in the models 
explaining Gamma Ray Bursts, Active Galactic Nuclei etc. 
Progress in numerical simulations over the years, however, have prompted 
the BZ mechanism from being merely 
speculative or theoretical to being close to the working model in nature 
\cite{MG2004,Palenzuela2010,Mosta2010}.
Simulations of two merging black-holes showed emergence of Poynting flux
jets with luminosities and energy scales as predicted by the BZ mechanisms 
\cite{Palenzuela2010,Neilsen2011}.
The energy output in the presence of surrounding plasma was an order of 
magnitude larger than in the electrovacuum case, demonstrating that 
the plasma facilitates efficient tapping of the black-hole energy.

Ref.~\refcite{Palenzuela2010} carried out a simulation of the interaction of a binary 
BH system, consisting of two non-spinning BHs of equal masses ($M=10^{8} M_{\odot}$),
with the surrounding magnetic field and conducting plasma. As the GW emission
drained off energy and momentum the BHs collided and merged, settling into 
a larger rotating Kerr BH with spin $a = Jc/GM^{2} \approx 0.67$ where $J$
is the BH's angular momentum. The merging BHs gave rise to a flare of 
Poynting flux ($L\approx 10^{43} ~erg/s$ lasting over 5 hrs) coincident with 
the peak of GW emission.

The electromagnetic energy released in the form of Poynting flux during 
the BH-BH merger could be expressed as a sum of spin and speed 
components \cite{Neilsen2011}: 
\begin{eqnarray}
L_{total} 	& = & L_{spin} + L_{speed}\\
		& = & \left[0.87\left(\frac{a}{0.6} \right)^{2} + 127 \beta^{2}\right] 
		M^{2}_{8} ~B^{2}_{4} ~L_{43} \rm ~erg ~s^{-1}
\end{eqnarray}
As can be seen, the non-spinning contribution to the total luminosity exceeds
that due to spin for $\beta \gtrsim 0.1$ while $\beta \gg 0.1$ in the last 
stages of merger and as a result the luminosity rises sharply.
Some of this may emerge as a synchrotron emission, as suggested 
by Ref.~\refcite{Palenzuela2010} in the form of a flare and the rest could be 
dissipated at large distances and could power the adiabatic expansion 
of the shock front traveling at relativistic to mildly relativistic speeds
as the total luminosity could easily exceed Eddington luminosity of the system.
The luminosity could be expressed in the units of Eddington luminosity 
$L_{Edd}$ i.e.
\begin{equation}
\frac{L_{total}}{L_{Edd}} \approx 0.1 \times B^{2}_{4} M_{8}
\end{equation}
where we have used $\beta=0.8$.
Conservatively, Ref.~\refcite{Palenzuela2010} chose the magnetic field 
such that $L \sim 0.002 L_{Edd}$ at merger. The limiting magnetic 
field required to reach this limit is $B=6\times 10^{4} \sqrt{M_{8}} ~G$.
The inner accretion disk could develop significantly high magnetic field, however, 
as shown by Ref.~\refcite{Pessah2006,Begelman2007}.
For $B_{4} = 10$ one gets $L_{total} \approx 7 \times L_{Edd}$
for the merger of BHs with $M = 10^{8} M_{\odot}$.
As shown by Ref.~\refcite{Pessah2006,Begelman2007} and Ref.~\refcite{Kaplan2011}
magnetorotational instability at the inner edge of the accretion disk could amplify 
magnetic fields to even larger values dependent on the mass of the central object
$B\approx 10^{6} M^{-7/20}_{8} ~G$. With this magnetic field the whole range of 
BHs masses, from stellar to supermassive, result in super-Eddington luminosities
having $L_{total}/L_{Edd}$ ranging from unity to several hundreds.

From their simulations Ref.\refcite{Palenzuela2010} estimate the total electromagnetic 
energy radiated during the merger to be $E \approx 2\times10^{48}$ erg.
As the BH-BH spiral in and come closer, their speed increases sharply 
and so does the radiated luminosity. The total energy could then be estimated
by integrating over the merger duration :
\begin{equation}
E \approx 127 \int^{t(\beta_{max})}_{t(\beta_{1})} \beta^{2} M^{2}_{8} B^{2}_{4} L_{43} dt
\end{equation}
where the integration starts from moderate velocity $\beta_{1}$ to the terminal 
velocity $\beta_{max}$. Following Ref.~\refcite{O'shaughnessy2011} the evolution 
of circular velocity is given by $d\beta/dt = A \beta^{9}$
where A is the proportionality constant $A = (96/15) c^{3} \eta/G M$ and 
$\eta = m_{1} m_{2} /M$ is the mass ratio of merging black-holes of masses 
$m_{1}$ \& $m_{2}$ with the total mass being $M = m_{1} + m_{2}$.

Knowledge of actual magnetic fields in the circum-binary disks threading the BHs
appears to be a limiting factor in further understanding of these mergers.
Therefore instead of assuming optimistically high energy release, we will follow 
Ref.~\refcite{Palenzuela2010} and Ref.~\refcite{Kaplan2011} such that $L \approx \epsilon_{Edd} L_{Edd}$
and absorb our ignorance about the magnetic fields and details of emission mechanism 
in the parameter $\epsilon_{Edd}$. The total energy release could then be expressed as
\begin{equation}
E \approx 2 \times 10^{50} \epsilon_{Edd} ~\eta^{-1} M^{2}_{8}
\label{eqn:BHenergy}
\end{equation}
where we have assumed moderate values for $\beta_{1} = 0.3$ and $\beta_{max} = 0.4$.

Thus, the most massive BH-BH mergers could be comparable to or more energetic than
the NS-NS mergers. Other characteristics, such 
as the rise time and brightness which are important parameters for carrying 
out  searches of the associated electromagnetic transients will also depend 
on the merger masses and energy outputs.

Ref.~\refcite{Kaplan2011} has shown that such a flare of synchrotron radiation 
could be seen from cosmological distances ($z\sim 2$). 
Since $\tau_{flare} \propto M_{BH}$ stellar mass BHs would produce flares
which will be both shorter and less intense than those produced by SMBHs.
Therefore recovering most of the stellar mass mergers would require 
a survey of rapid cadence. As shown by Ref.~\refcite{Kaplan2011} a radio survey 
operated around $GHz$ frequencies with a cadence of 10s would be able 
to see $\approx 24$ mergers per year most of which would be due to stellar 
mass mergers. A slow cadence will not see stellar mass mergers but will 
see about 3 mergers per year produced by SMBHs.

It is possible that a comparable amount of energy is released in the expansion 
of the shock wave into the surrounding material. This shock wave can shine 
as the ``afterglow'' transient in the days after the flare peaked during 
the merger and GW emission. 
The amount of energy will depend on the masses of the merging black holes 
as per the Equation~\ref{eqn:BHenergy} above and can therefore range from 
$10^{36}$ erg for $M = 10~M_{\odot}$ to $10^{51} erg$ for $10^{8}~M_{\odot}$ 
BHs.

The merging masses will also determine the timescales over which 
the transient will be shining. As can be seen from Eqn. 16
the shock will start to decelerate at $t_{dec} \approx 45 E^{1/3}_{49}$ days
around which the EM transient will be at its peak brightness; effectively
between a few minutes for the mergers of stellar mass BHs to a few months 
for $M = 10^{8}~M_{\odot}$, respectively.

Peak brightness of the transient at a distance of $100 Mpc$ would be 
$F_{\nu,peak} \approx 0.5 E_{49}$ mJy implying that mergers
of stellar mass BHs might be too faint to detect but those of super-massive BHs 
could be detectable out to about $Gpc$ with the present sensitivity of 
the radio telescopes.

\section{Identifying radio ``afterglow'' of compact mergers}
As discussed above the expected detection rate of mergers of compact objects 
events and SNe will be comparable (a few events per year). 
Evolution of their light curves would be similar and therefore 
distinguishing them from each other would be important. 
Multi-wavelength observations, especially optical spectroscopy will be useful 
to identify compact mergers from other similar transients.

The most important early distinction could be made by using optical
observations. The compact merger events are expected to be faint
in the optical unless their optical light is powered by a process
other than synchrotron radiation, such as the decay of radio active
elements synthesized in the explosion. Deep limits on the optical
brightness of a transient should be able to rule out the presence of
SNe. Furthermore, optical spectra of SNe have been studied in detail
and its evolution in time is also very well understood (e.g. for a review see 
Ref.~\refcite{1997ARA&A..35..309F}. Also see Ref. \refcite{2006ARA&A..44..507W} 
and references therein.) Optical light curves of SN also have characteristic 
shapes \cite{1996sunu.book.....A}. Together these
features should be able to identify a SNe from a compact merger event as early as
within a week to a month from the discovery. 
Over longer time-scales mutli-wavelength radio observations of such transients
can play crucial role in making the distinctions by carrying out detailed 
calorimetry of explosions. Multi-band colorimetry has been carried out in several
GRB afterglows and SNe \cite{Berger2003,Soderberg2006,AvdH2008,AvdH2011}.
Compact mergers are expected to release up to $10^{49}$\,erg as kinetic energy
as opposed to SNe whose typical energy budgets are $10^{50}-10^{51}$\,erg.

\section{Conclusion}
In the near future several advanced gravitational wave observatories will become 
operational and are expected to detect first GWs. This will open an entirely new
window on the transients and on the Universe. GWs will become a new 
probe of the Universe and our understanding of almost every aspect of the Universe,
from stellar evolution, galaxy formation to the early universe,
will be revolutionized. To substantially exploit the benefits
of GW discoveries, however, it will be important to discover EM counterparts 
of GW sources. 

Among other sources of gravitational waves, mergers of 
compact objects such as NS-NS, NS-BH and BH-BH are considered prime 
sources of GWs. Many of these mergers could be explosive and would 
drive blast waves,  relativistic or non-relativistic, into the surrounding medium
and shine as bright radio transients. We have shown that these events would
produce transients bright enough to be detected from up to 300 Mpc
and about half a dozen should be visible in the entire sky at any given time 
to the radio telescopes operating at GHz frequencies.

The arrival of GW detectors online is most opportune as several modern technology
radio telescopes will soon be operational across the globe : already operational 
Merchison Widefield Array\footnote{http://www.mwatelescope.org/} 
and Australian SKA Pathfinder\footnote{http://www.atnf.csiro.au/projects/mira/} 
(to be operational by 2013) in Australia, MeerKat\footnote{http://www.ska.ac.za/meerkat/index.php}
 (by 2018) in SouthAfrica, Low Frequency Array \footnote{http://www.lofar.org/}
in the Netherlands and Europe is already operational while Apertif
\footnote{http://www.astron.nl/general/apertif/apertif} in the Netheralands 
will be operational by the end of 2012, and the Square Kilometer Array
\footnote{http://www.ska.gov.au/Pages/default.aspx} (by 2024) from 
South Africa and Australia. Besides having significantly better sensitivity
they also have wide field imaging capabilities which makes continuous 
scanning of the sky possible. Wide and moderately deep sky surveys carried 
out by these telescopes should uncover at least a few transients each year.
With comparatively better localization of the transients made available by these 
telescopes spatial coincidence of the transients should reveal their asociation 
with the GW sources detected by GW observatories.

The GHz frequency radio sky survey appears to be a promising method 
to discover transients associated with the outflows
generated by merges of compact objects. Blind radio surveys should be able to detect
about a few of such sources at any given time in the sky.
Mergers of NS-NS binaries should produce transients with total energy
output of about $10^{49}$ erg. The total energy output of BH-BH binary 
mergers is less uncertain but is likely to be lower by several orders of magnitude
for stellar mass BH mergers compared to those of the super-massive BHs.
As a result afterglow like emission in the stellar mass BH mergers would 
be very faint to be detectable for the current or near future radio telescopes.

The detection rate of compact object mergers will be comparable to that of
the radio supernovae. Besides, the light curve evolution of radio supernovae 
and afterglows from compact object mergers would look alike. 
Therefore distinguishing the two classes from each other observationally 
will be important. This could be done easily at early times by optical spectroscopy.
Detailed calorimetry could be carried out by long term mutli-band follow up
which could also aid in distinguishing the compact object mergers from radio supernovae.

\section{Acknowledgement}
A.P.K. gratefully acknowledges organizers of the ASTROD-5 meeting and 
Raman Research Institute for invitation, support and hospitality. 
We thank the anonymous referee for a prompt and helpful review which 
improved this article.
A.P.K. and D.L.K. were partially supported by NSF award AST-1008353.
\bibliography{CompactMerger}

\end{document}